# Magnetic structure in the two-dimensional van der Waals ferromagnet $Fe_3GaTe_2$

**Po-Chun Chang[a, b*]. Sabreen Hammouda[a]. Yung-Hsiang Tung[a]. Yishui Zhou[a]. Iurii Kibalin[c]. Bachir Ouladdiaf[d]. Chao-Hung Du[b]. and Yixi Su[a*]**

[a] Jülich Centre for Neutron Science JCNS at Heinz Maier-Leibnitz Zentrum (MLZ), Forschungszentrum Jülich GmbH, Lichtenbergstraße 1, Garching D-85747, Germany

[b] Department of Physics, Tamkung University, New Taipei City 251301, Taiwan

[c] European Spallation Source ERIC, P.O. Box 176, SE-221 00 Lund, Sweden

[d] Institut Laue-Langevin, 71 avenue des Martyrs, Grenoble, 38042, France

Correspondence e-mail: chang155212@gmail.com (Po-Chun Chang), y.su@fz-juelich.de (Yixi Su)

**Synopsis** High-quality single crystals of $Fe_3GaTe_2$ were grown using the chemical vapor transport method, allowing for a precise determination of their crystal and magnetic structures. A shortening of the $Fe^i$–$Fe^i$ distance enhances the Fe–Fe exchange interaction, providing a microscopic origin for its higher Curie temperature compared with $Fe_3GeTe_2$.

**Abstract** High-quality single crystals of the two-dimensional van der Waals ferromagnet $Fe_3GaTe_2$ (FGaT) were successfully grown using the chemical vapor transport method, which effectively reduced surface impurities compared with conventional self-flux growth. Structural and magnetic characterizations were performed using single-crystal X-ray and neutron diffraction. The results confirm that FGaT crystallizes in the hexagonal $P6_3/mmc$ structure, with Fe occupying two inequivalent sites ($Fe^i$ and $Fe^{ii}$), where the magnetic moment of $Fe^i$ [1.9(2) $\mu_B$] is larger than that of $Fe^{ii}$ [1.4(6) $\mu_B$]. The magnetic easy axis is oriented along the $c$ axis and the Curie temperature ($T_C$) is approximately 355–360 K. Compared with $Fe_3GeTe_2$ (FGT), FGaT exhibits a slightly expanded $a$ axis and a contracted $c$ axis, resulting in a reduction in the $Fe^i$–$Fe^i$ interatomic distance along the $c$ axis. This pronounced contraction could strengthen the Fe–Fe exchange interaction, which is believed to be the key factor responsible for the significantly higher $T_C$ in FGaT relative to FGT.

**Keywords: $Fe_3GaTe_2$; van der Waals ferromagnet; single-crystal neutron diffraction; magnetic structure.**

## 1. Introduction

In the field of spintronics, magnetic two-dimensional van der Waals (2D-vdW) materials have garnered considerable attention due to their potential applications, particularly in magnetic tunnel junctions and spin current transport, owing to their unique structures and magnetic anisotropy. As a result, they have become a frontier in contemporary condensed matter physics and materials science research. FGaT is a new type of two-dimensional van der Waals ferromagnet [1] with an extremely high $T_C$ of approximately 350-380 K, which has attracted numerous research teams over the past two years. Recent studies have explored various approaches to utilizing FGaT in spintronic devices, including achieving magnetic switching through current-induced spin-orbit torque [2-4], regulating the $T_C$ by applying a perpendicular electric field [5], and manipulating interfacial electronic properties by forming heterostructures with different two-dimensional materials [6]. In addition, it was reported that FGaT forms iron atom defects easily, which produce spatial inversion symmetry breaking, leading to Dzyaloshinskii-Moriya interactions, and ultimately forming topologically protected skyrmions under external magnetic field induction [7, 8], which has great potential for information storage and processing applications. Furthermore, because FGaT had a $T_C$ above room temperature, the development of room-temperature magnetic 2D-vdW materials with novel magnetism through Co or Ni doping is studied [9, 10]. These diverse applications were based on the combination of perpendicular magnetic anisotropy and high $T_C$ in FGaT.

The atomic structure and magnetic properties of FGaT exhibit similarities to those of another related, but better-studied, 2D-vdW ferromagnet, FGT. However, the $T_C$ values of the two materials mentioned above are significantly different, with the $T_C$ of FGT being only 170-220 K, depending on the material defects [11]. Some studies try to explain the origin of its high $T_C$ through first-principles calculations, focusing on the density of states (DOS) near the Fermi level [12, 13]. Other studies have shown that changes in the lattice constant can cause internal stress or atomic displacement, thereby altering the exchange interaction and contributing to the enhancement of $T_C$ [14]. All of these factors could lead to a higher $T_C$ for FGaT than for FGT. However, some reports suggest that $T_C$ might actually decrease when the FGT crystal structure is compressed under high pressure [15, 16]. In addition, some studies showed that the intercalated Fe between van der Waals layers leads to enhanced magnetic interlayer coupling, which may be one of the reasons for the increase in $T_C$ [17, 18]. However, this phenomenon may also be caused by high-temperature pre-annealing [19] and therefore may not be present in all experiments.

The above studies all highlight the cause of the high $T_C$ of FGaT, which remains a topic worthy of in-depth exploration. In this study, we grew single crystals of FGaT using CVT to avoid impurities in the sample. SC-XRD and single-crystal neutron diffraction were employed to investigate the differences in the atomic and magnetic structures of FGaT compared to FGT, with the aim of identifying the fundamental cause of the difference in $T_C$ and contributing to the study of new room-temperature ferromagnetic 2D-vdW materials.

#### 2. Experimental

High-quality single-crystal samples were produced by using the CVT method. Lump Fe (99.99%), lump Ga (99.99%), and columnar Te (99.99%) were placed in a 14cm-long quartz tube in an Ar-filled glove box with an atomic molar ratio of 3:1:2. A total of 5 grams of raw material was added to the quartz tube, with iodine as a transport agent at approximately 2 mg $cm^{-3}$ [20-21]. The tube was sealed at an internal pressure of $1x10^{-3}$ mbar. To prevent iodine vaporization during the sealing process, the end of the tube was immersed in water for cooling. The furnace, with a hot zone of 1033 K and a cold zone of 983 K, was set for 168 hours. FGaT single crystals were obtained after cooling naturally to room temperature. The sample quality was analyzed using X-ray diffraction (XRD) data collected on a Bruker D2 PHASER diffractometer, which utilized Cu K$\alpha$ radiation ($\lambda$ = 1.5406 Å) to examine the out-of-plane structural signals of the two-dimensional material. The magnetic susceptibility measurements were performed using a Quantum Design PPMS DynaCool system. For the $\chi$–$T$ measurements, the Zero Field Cooling (ZFC) data were collected upon warming, while the Field Cooling (FC) data were collected upon cooling. A DC magnetic field was applied at 2 K to measure the magnetization, which was then used to estimate the average magnetic moment per unit cell. SC-XRD data were collected using a Rigaku XtaLAB Synergy-S diffractometer ($\lambda$ = 0.71073 Å), analyzed with *CrysAlisPro*, and refined with *JANA2006* [22] and *FullProf* [23] for atomic positions and occupancies. Neutron diffraction data were collected with a wavelength of 1.24 Å on the D10+ instrument at Institut Laue–Langevin (ILL), and the magnetic structure was refined using *FullProf* [23].

#### 3. Results and Discussion

High-quality single crystals of FGaT were grown using CVT, exhibiting a hexagonal layered crystal structure of space group $P6_3/mmc$. Each van der Waals layer was formed by Te on the outside, with $Fe_3Ga$ layers sandwiched in between. Fe exists in two positions within the unit cell, denote $Fe^{i}$ and $Fe^{ii}$. The van der Waals layers were stacked with their normal parallel to the c axis, similar to FGT [Fig. 1(*a*)]. Fig. 1(*b*) shows the XRD diffraction pattern of the freshly grown sample in the out-of-plane direction. All diffraction signals are from the (00*l*) planes of the two-dimensional material, with no extra diffraction, indicating a lack of surface impurities. On the other hand, if the self-flux method, as used in most reports, is employed to grow crystals, it is impossible to obtain crystals large enough and with low impurity content for a single-crystal neutron diffraction experiment. We initially attempted to grow crystals using the self-flux method in the molar ratio of Fe:Ga:Te = 1:1:2. However, the XRD diffraction pattern along the *c* axis revealed an additional diffraction peak at 26.2° ($2\theta$) besides the reflections from the (00*l*) planes. This was probably from a $Ga_2Te_3$ alloy that had on the surface of the FGaT crystal, confirmed by energy dispersive spectrometer (EDS). In contrast, EDS analysis of FGaT crystals grown via CVT showed a uniform elemental ratio of Fe:Ga:Te = 3:1:2 throughout the crystal. Therefore, both XRD along the *c* axis and EDS analysis confirmed that the FGaT crystals grown using the CVT method were free of surface impurities.

Fig. 1(*c*) shows a photograph of the crystals as grown, while Fig. 1(*d*) shows a photograph of the sample used in this neutron single-crystal diffraction experiment and the corresponding X-ray Laue image. The temperature-dependent magnetic susceptibility in different directions, and the hysteresis loops measured at 2 K and 300 K parallel to the *c* axis, are shown in Figs. 1(e) and (f), respectively. For the $\chi$–$T$ measurements, the ZFC curve was recorded during warming, while the FC curve was recorded during cooling. The small offset between the ZFC and FC curves is attributed to a slight difference in reaching thermal equilibrium of the sample between these two measurement protocols. The magnetic susceptibility measurements in different directions confirm that the magnetic easy axis was along the *c* axis, and the $T_C$ was estimated to be approximately 355–360 K using the Curie–Weiss law. The remanence of the hysteresis loop at 300 K was almost zero because the striped magnetic domains formed without the application of a magnetic field [7].

Due to the high sample purity, it was possible to calculate the average magnetic moment per Fe atom within the unit cell with the molar mass. Based on the magnetic moment values obtained in the saturation magnetic field region at 2 K, the average magnetic moment per Fe atom at 2 K was estimated to be approximately 1.76 $\mu_B$.

SC-XRD measurements were used to analyze the structural details of FGaT. The lattice parameters of FGaT were determined during the SC-XRD experiment using a Rigaku XtaLAB Synergy-S diffractometer and analyzed with *CrysAlisPro*. The resulting values are *a* = 4.0794 Å, *c* = 16.09040(10) Å. A total of 411 reflections were used in the refinement. Previous studies have suggested that Fe atoms adjacent to Ga may undergo slight displacements along the c-axis due to structural vacancies, leading to additional diffraction signals at (*hhl*) (where $l = 2n + 1$). These reflections are forbidden in the centrosymmetric space group $P6_3/mmc$ and therefore require the non-centrosymmetric space group $P3m1$ for structural description [7-8]. However, in our single-crystal diffraction measurements [Fig. 2(*a*)], no (*h, h, l*) reflections with $l = 2n + 1$ were detected, indicating the absence of $P3m1$ symmetry. Thus, the crystal structure of our FGaT sample is best described by the $P6_3/mmc$ space group. Furthermore, the measured lattice parameters reveal that the *a* axis of FGaT is slightly longer, whereas the *c* axis is slightly shorter than those of FGT [24].

The XRD refinement results of the nuclear structure are shown in Fig. 2(*c*). Based on the refinement results, we further analyzed the occupancy of Fe atoms at different positions. Since it was generally believed in the literature that Ga atoms have no obvious defects, this study only focuses on the defect of Fe [1]. Table 1 summarizes the refinement results of atomic positions and occupancy. Approximately 5% vacancy at the $Fe^{i}$ and 16% vacancy at the $Fe^{ii}$ were confirmed.

To analyze the magnetic structure of $Fe_3GaTe_2$, neutron single-crystal diffraction experiments were conducted on beamline D10+ at the ILL. To collect sufficient diffraction signals, diffraction peaks were collected from a 10 mg single crystal using a four-circle configuration of D10+ at 2 K and at room temperature. However, using the four-circle configuration comes at the cost of not being able to heat the sample above 360 K. Therefore, the subsequent refinement of the atomic structure and related information is based on data obtained from XRD refinement.

The crystal quality was initially assessed by recording Laue diffraction patterns [Fig. 1(*d*)], which confirmed that the sample is a single domain and possesses hexagonal symmetry. The diffraction intensities of (002) and $(1\bar{2}0)$ at 300 K and 2 K are compared in Fig. 3(*a*), confirming that the intensity of $(1\bar{2}0)$ increased as the temperature is decreased, but the intensity of (002) does not, indicating that the components of the magnetic moment are indeed aligned along the *c* axis and have no components in the *ab* plane.

The diffraction signal of (00*l*) at 2 K was used to confirm the symmetry of the structure. Fig. 2b shows the diffraction signal of (00*l*) ($l = 2n + 1$) and (006) at 2 K for comparison. Although (006) was a very weak structural diffraction peak under neutron diffraction, no diffraction peaks of the odd-numbered (00*l*) can be observed, confirming that no $Fe^{ii}$ shift is observed under neutron diffraction.

Figs. 3(*c*) and (*d*) show the refinement results of the atomic and magnetic structures at 2 K. Based on the previously published X-ray Magnetic Circular Dichroism (XMCD) studies by Zha and co-workers, Fe has been identified as the dominant source of magnetism in FGaT, while the contributions from Ga and Te are negligible [25]. In addition, the magnetic susceptibility measurements indicate a ferromagnetic alignment of the magnetic moments along the $c$-axis. On this basis, magnetic models with moments oriented along the $a$ or $b$ directions, as well as antiferromagnetic arrangements along the $c$-axis, can be excluded.

To build the magnetic structure model, the *MAXMAGN* tool from the Bilbao Crystallographic Server was used [26]. After excluding those models that do not match experimental observations or which fail to converge after refinement, the magnetic structure was refined with the magnetic space group $P6_3/mm'c'$ (No. 194.270), in which the magnetic moments of $Fe^{i}$ and $Fe^{ii}$ were parallel to each other and aligned along the $c$-axis. In total, 57 reflections were used in the refinement. Because the structural parameters were fixed to the XRD results, only the scale factors and the magnetic moments on the Fe sites could be refined in the single-crystal neutron diffraction analysis. The average magnetic moment, 1.76 $\mu_B$ per Fe atom, based on our magnetic susceptibility measurement at 2 K, was considered in magnetic structure refinement. In the final refinement, the scale factor was fixed so that the average magnetic moment of the two Fe sites, taking site occupancies into account, was consistent with the value obtained from magnetic susceptibility measurements. Under this constraint, the refined magnetic moments are $Fe^{i}$ = 1.9 (2) $\mu_B$ and $Fe^{ii}$ = 1.4 (6) $\mu_B$. Therefore, the magnetic moments on the two Fe sites were not refined fully independently in the final model, and this constraint was applied to ensure consistency with the bulk magnetic measurements.

The magnetic moment of FGaT revealed by the single-crystal neutron diffraction is similar to that of FGT, which was determined on neutron powder diffraction analysis [24]. Both results were compared in Table 2. Many first-principles theoretical calculations have also obtained values close to these results [10, 12, 25], indicating that two magnetic structures are generally similar. However, the $T_C$ of FGaT was much higher than FGT, indicating that the key factor affecting $T_C$ may not come from the magnetic structure, but rather be related to the electronic structure of each atom in the material, the strength of the exchange interaction, and the difference in orbital overlap caused by the changes in the lattice constant.

Other possible explanations for the high $T_C$ of FGaT have also been studied. For example, each Fe site in $Fe_5GeTe_2$ had a different occupancy due to the thermal annealing processes used [27], and a similar situation occurs in FGaT. Previous studies have shown that incorporating a small number of Fe atoms into the van der Waals gaps, along with antiferromagnetic coupling, yields simulated results that more closely match the neutron powder diffraction data [18]. These interlayer Fe atoms were thus suggested as a possible origin of the enhanced $T_C$ [18]. However, it was subsequently reported that the formation of such intercalated Fe can be induced by annealing. Even in the absence of Fe intercalated atoms, the $T_C$ of FGaT was still much higher than room temperature [19]. Therefore, intercalated Fe should not be the main reason for the high $T_C$ phenomenon. In addition, since our sample was not repeatedly annealed after growth, the case of Fe intercalation was not a concern.

In the FGT system, studies used density functional theory (DFT) to confirm clearly that its magnetism comes from local magnetic moments and itinerant electrons [28, 29], so it can be considered a typical Hund metal. Although FGaT also exhibited the coexistence of localization and itinerant characters, its overall magnetism is more clearly dominated by local magnetic moments contributed by $Fe^{i}$ [13, 30]. Lee and co-workers further demonstrated that the reduction in the Fe–Fe interatomic distance enhances the overlap between Fe 3d orbitals, leading to a substantial increase in the nearest-neighbor Heisenberg exchange constant $J_1$ [14]. This strengthened exchange interaction is considered to be a key factor in the higher $T_C$ observed in FGaT relative to FGT.

Based on the results of SC-XRD refinement, a comparative analysis was conducted between the crystal structures of FGT and FGaT. The refined interatomic distances of FGaT are summarized in Table 3, while the corresponding parameters for FGT can be found in previous reports on its atom and magnetic structure [24]. The refinement results show that FGaT expands slightly along the *a* axis and contracts slightly along the *c* axis, resulting in slight changes in the interatomic spacings compared with FGT. Notably, the $Fe^{i}$-$Fe^{i}$ distance along the *c* axis in FGaT exhibits the largest variation compared to that in FGT. Such a pronounced contraction is expected to strengthen the $Fe^{i}$–$Fe^{i}$ exchange interaction and enhance the nearest-neighbor Heisenberg exchange-coupling interaction, thereby increasing the thermal energy required to destabilize the ordered magnetic state.

To place this structural difference in a clear physical context, a comparison between FGT and FGaT was performed using reported experimental and theoretical results. In the work of Ghosh *et al.* on FGT, first-principles (DFT) calculations via GGA+DMFT showed that for an $Fe^{i}$–$Fe^{i}$ distance of 2.47 Å along the *c* axis, the nearest-neighbor exchange interaction parameter $J_1$ is approximately 37 meV [31]. In contrast, Lee *et al.* reported that for FGaT a shorter $Fe^{i}$–$Fe^{i}$ distance of 2.409 Å along the *c*-axis leads to an increased exchange parameter $J_1$ = 74.83 meV [14]. Experimentally, Verchenko and co-workers reported an $Fe^{i}$–$Fe^{i}$ distance of 2.602 Å for FGT [24], whereas in our present work, the $Fe^{i}$–$Fe^{i}$ distance along the $c$-axis is refined to be 2.479 Å for FGaT. Ghosh *et al.* further showed that when only GGA is used (without DMFT), the exchange parameter $J_1$ in FGT increases to 65.32 meV, yielding an estimated $T_C$ of approximately 470 K [31]. These results indicate that a reduction in $Fe^{i}$–$Fe^{i}$ distance can indeed lead to an increase in the exchange parameters and consequently to an enhancement of $T_C$. However, the quantitative relationship between distance contraction and the exact change in exchange parameter remains highly model dependent. For example, Li and co-workers reported an $Fe^{i}$–$Fe^{i}$ distance of 2.37 Å with an exchange parameter $J_1$ = 57.18 meV, illustrating that substantial variations may exist among different theoretical studies. Based on these comparisons, we conclude that the $Fe^{i}$–$Fe^{i}$ distance in FGaT is shorter than in FGT by approximately 0.12 Å [24], and that this reduction corresponds, according to various DFT calculations, to an increase in the exchange parameter $J_1$ in the order of roughly 20–40 meV [14, 32]. Such a substantial increase is sufficient to account for an enhancement of $T_C$ from approximately 220–230 K to 350–380 K, consistent with the experimental observations.

Regarding the relative roles of the Ga site and the $Fe^{ii}$ site in influencing the interlayer $Fe^{i}$–$Fe^{i}$ distance, this issue can be addressed by comparing the structural differences between FGaT and FGT, as well as the effect of $Fe^{ii}$ vacancies in FGT. For FGT, the lattice parameters are $a$ = 4.00848(2) Å and $c$ = 16.3307(1) Å [24]. In contrast, our SC-XRD measurements on FGaT yield $a$ = 4.0794 Å and $c$ = 16.09040(10) Å, in excellent agreement with previously reported values ($a$ = 4.09(2) Å, $c$ = 16.07(2) Å) [14]. This comparison suggests that substitution of Ge by Ga leads primarily to a shortening of the $c$ axis and a slight expansion of the $a$ axis, resulting in a reduced $Fe^{i}$–$Fe^{i}$ distance. In contrast, May and co-workers showed that in $Fe_{3-x}GeTe_2$, increasing $Fe^{ii}$ vacancy concentration led to a decrease in the lattice parameter $a$ and an increase in $c$, accompanied by a reduction of $T_C$ [11]. Similarly, Bao and coworkers reported that defective FGT exhibits a reduced $T_C$ down to approximately 160 K [28]. These observations indicate that $Fe^{ii}$ vacancies do not promote a contraction of the $Fe^{i}$–$Fe^{i}$ interlayer distance. Instead, they tend to have the opposite effect. Therefore, the $Fe^{i}$–$Fe^{i}$ distance reduction observed in FGaT is most reasonably attributed to the Ga/Ge substitution rather than to $Fe^{ii}$ vacancies. Furthermore, the contraction of the $c$ axis would normally be expected to shorten the Te–$Fe^{i}$ distances. However, the excessive reduction of the $Fe^{i}$–$Fe^{i}$ spacing instead results in a slight elongation of the Te–$Fe^{i}$ bond. The influence of Te on the Fe electronic states remains largely unexplored and requires further investigation.

## 4. Conclusions

In this study, we have demonstrated that the CVT method enabled the growth of higher-quality FGaT single crystals compared with using the self-flux method, thereby benefiting single-crystal diffraction experiments. SC-XRD and neutron diffraction measurements have been employed to investigate the atomic and magnetic structures of FGaT, confirming that the magnetic moment of $Fe^{i}$ was larger than that of $Fe^{ii}$.

After replacing Ge with Ga, the vdW layered structure becomes slightly thinner and wider, and there is a significant shortening of the $Fe^{i}$–$Fe^{i}$ distance along the *c* axis. This structural modification provided important insight into the physical reason for the significantly higher $T_C$ observed in FGaT than in FGT.

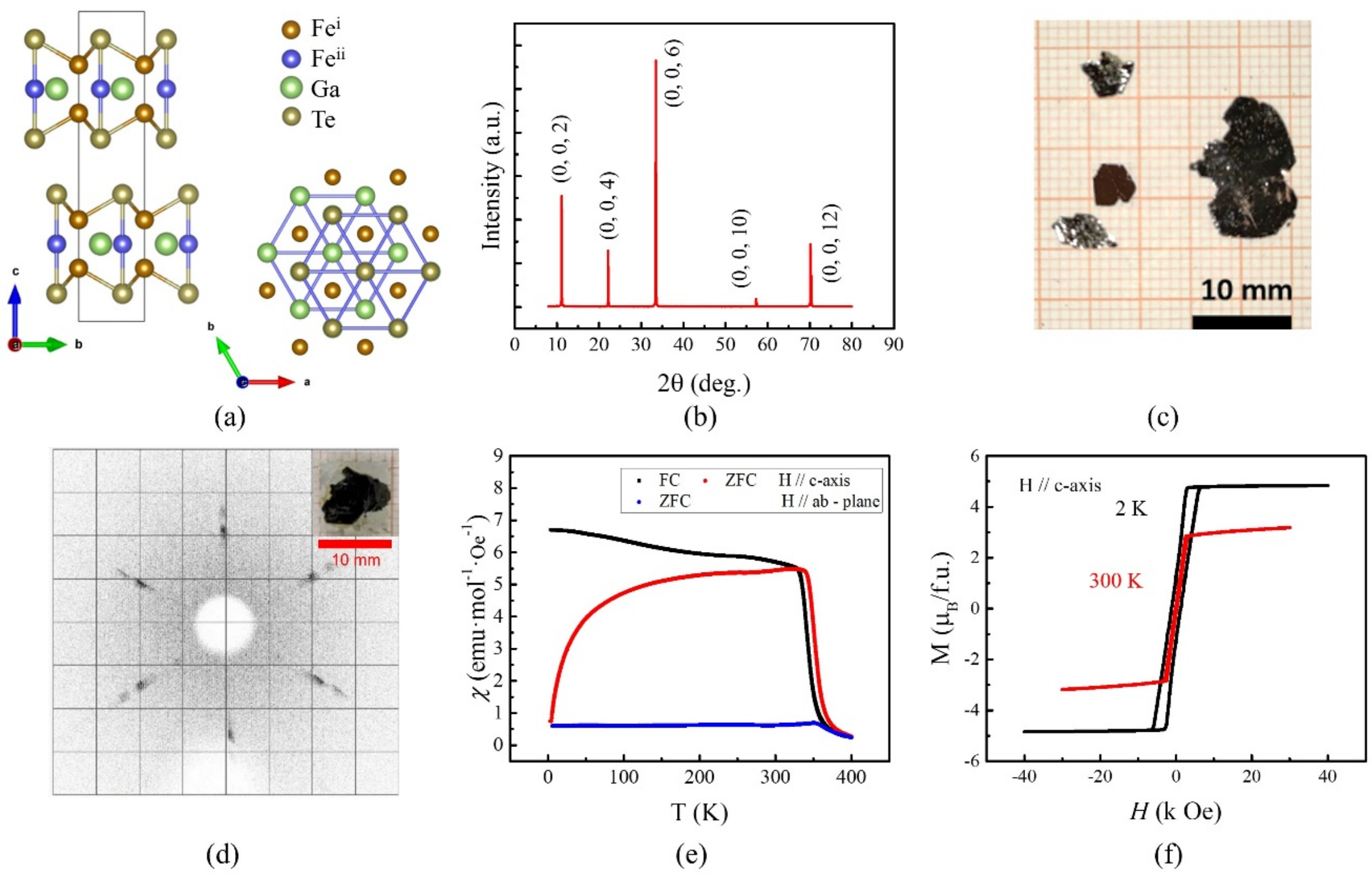

**Figure 1** (*a*) Schematic diagram of the side and top view of a van der Waals FGaT crystal. (*b*) XRD patterns along the normal face. (*c*) A photograph of the crystals as grown. (*d*) X-ray Laue image and (inset) photograph of the crystal used in neutron diffraction. (*e*) Temperature-dependent out-of-plane and in-plane magnetic susceptibility measurements under FC and ZFC conditions with 1000 Oe for bulk FGaT. (*f*) Out-of-plane magnetic hysteresis curves measured by VSM at 2 K and 300 K.

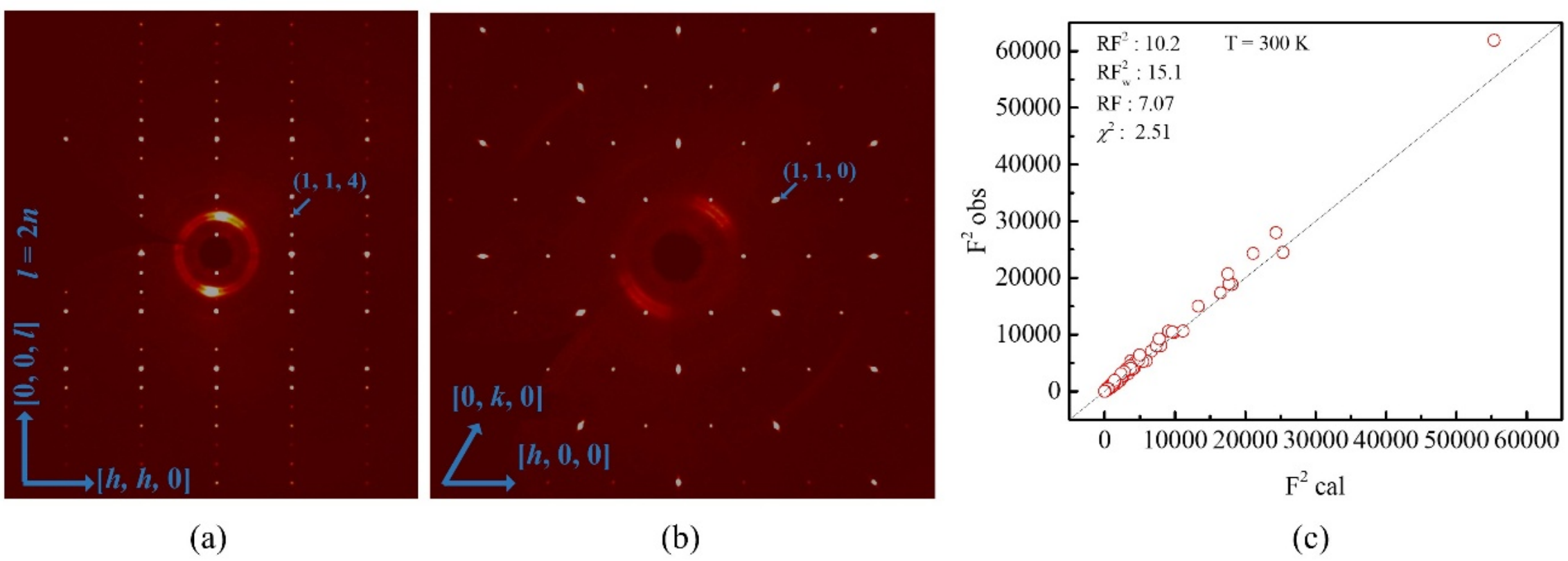

**Figure 2** (*a*) SC-XRD along [*hh*0] and [00*l*] direction. (*b*) SC-XRD along [*h*00] and [0*k*0] directions confirmed the crystal had hexagonal symmetry. (*c*) The refined nuclear structure result is in the space group $P6_3/mmc$.

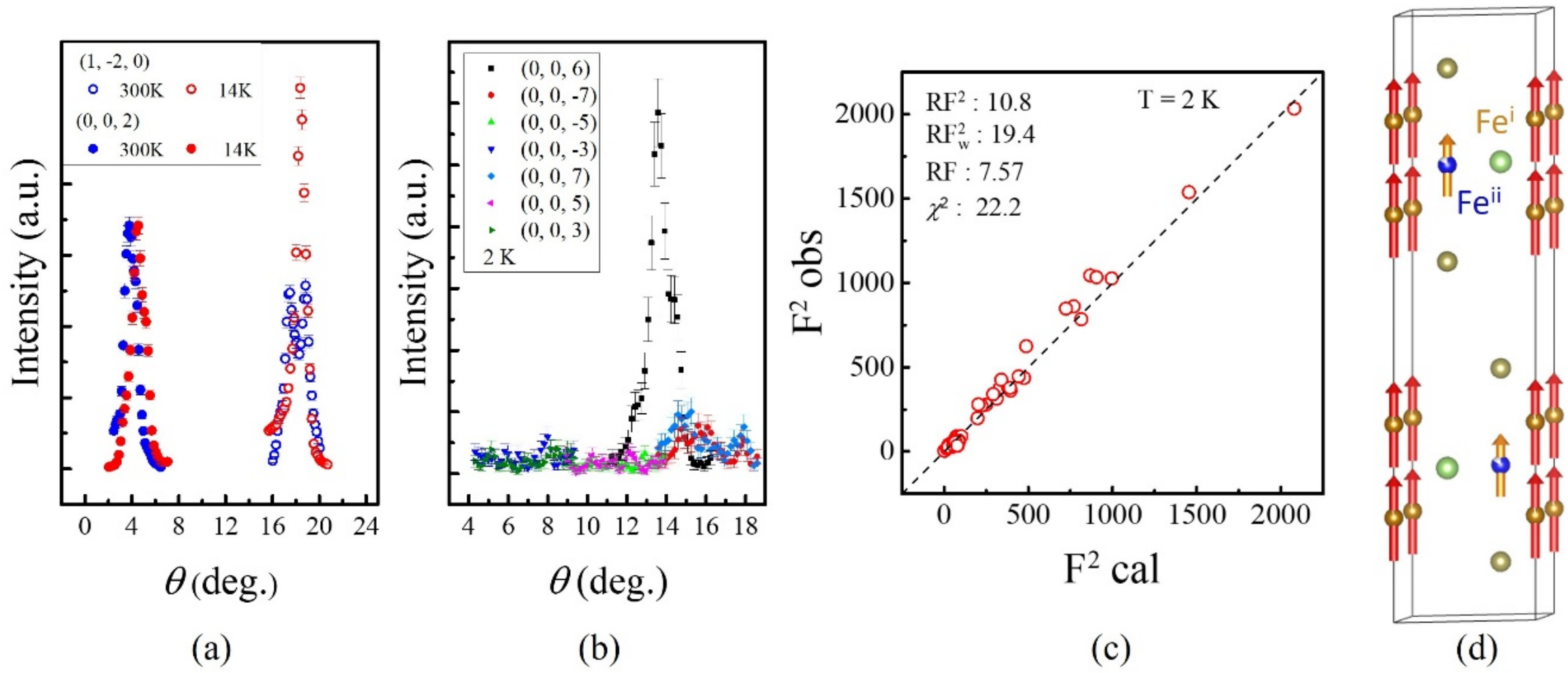

**Figure 3** (*a*) Temperature dependence of the (002) and ($1\bar{2}0$) reflection of single crystal neutron diffraction. (*b*) The diffraction signal of (00*l*) ($l = 2n + 1$) and (006) at 2 K. (*c*) The refinement result of the magnetic structure with the atomic structure at 2 K. (*d*) The magnetic structure refinement result with the space group $P6_3/mm'c'$. The magnetic moment of $Fe^{i}$ is larger than of the $Fe^{ii}$.

**Table 1** Atom parameters from SC-XRD of FGaT.

| Atom | Site | x | y | z | Occupancy | $B_{iso}$ (Å$^2$) |
|---|---|---|---|---|---|---|
| $Fe^{i}$ | 4*e* | 0 | 0 | 0.67296(8) | 0.95(1) | 0.839(31) |
| $Fe^{ii}$ | 2*d* | 2/3 | 1/3 | 3/4 | 0.841(16) | 0.732(46) |
| Ga | 2*c* | 1/3 | 2/3 | 3/4 | 1 | 1.665(35) |
| Te | 4*f* | 2/3 | 1/3 | 0.59053(4) | 1 | 1.032(16) |

**Table 2** Comparison of the magnetic moment of FGaT and FGT. The result of FGT are based on neutron powder diffraction [24].

| Parameter | FGaT | FGT |
|---|---|---|
| $M$ ($Fe^{i}$) ($\mu_B$) | 1.9(2) | 1.95(5) |
| $M$ ($Fe^{ii}$) ($\mu_B$) | 1.4(6) | 1.56(4) |

**Table 3** Interatomic distances of FGaT from SC-XRD, compared with reported values for FGT [24].

| Bond | | Distance in FGaT (Å) | Distance in FGT (Å) |
|---|---|---|---|
| Ga/Ge | $Fe^{i}$ | 2.6615(7) | 2.655 |
| | $Fe^{ii}$ | 2.35524(0) | 2.314 |
| $Fe^{i}$ | $Fe^{i}$ | 2.479(3) | 2.602 |
| | $Fe^{ii}$ | 2.6615(7) | 2.655 |
| | Ga | 2.6615(7) | 2.655 |
| | Te | 2.7030(8) | 2.611 |
| $Fe^{ii}$ | Ga | 2.35524(0) | 2.314 |
| | Te | 2.5659(7) | 2.613 |
| | $Fe^{i}$ | 2.6615(7) | 2.655 |

**Acknowledgements** The authors thank Markos Skoulatos and Bastian Veltel, of the Physics Lab at the Heinz Maier-Leibnitz Zentrum (MLZ), for their operational support in the physical property measurements. The single-crystal XRD measurements were performed at JCNS-MLZ, Garching. This study was financially sponsored by the National Science and Technology Council of Taiwan under grants NSTC 113-2811-M-032-002 and 113-2112-M-032-002. Open access funding enabled and organized by Projekt DEAL.

**Conflicts of interest** Po-Chun Chang mentioned that financial support was provided by National Science and Technology Council (NSTC) and Forschungszentrum Jülich GmbH. There are no other known conflicts of interest or personal relationships that could affect the work reported in this paper.

**Data availability** Neutron diffraction data from D10+ (ILL, France) are available at https://doi.ill.fr/10.5291/ILL-DATA.5-41-1243